\documentclass[12pt,a4]{article} 
\usepackage{amssymb,amsthm,url}
\usepackage{graphicx,hyperref}
\usepackage[english]{babel}

\begin{document}

\title{A New Sample of Gamma-Ray Emitting Jetted Active Galactic~Nuclei\\
Preliminary Results}

\author{
L. Foschini\footnote{Brera Astronomical Observatory, National Institute of Astrophysics (INAF), 23807 Merate, Italy.}, \and M. L. Lister\footnote{Department of Physics and Astronomy, Purdue University, West Lafayette, IN 47907, USA.}, \and S. Ant\'on\footnote{CFisUC, Departamento de F\'{i}sica, Universidade de Coimbra, 3004-516 Coimbra, Portugal.}, \and M. Berton\footnote{European Southern Observatory (ESO), Santiago de Chile 19001, Chile.}~\footnote{Finnish Centre for Astronomy with ESO (FINCA), University of Turku, 20014 Turku, Finland.}~\footnote{Mets\"{a}hovi Radio Observatory, Aalto University, 02540 Kylm\"{a}l\"{a}, Finland.}, \and S. Ciroi\footnote{Department of Physics and Astronomy, University of Padova, 35122 Padova, Italy.}, \and M.~J.~M. March\~{a}\footnote{Physics and Astronomy Department, University College London, London WC1E6BT, UK.}, \and M. Tornikoski$^{\parallel}$, \and E. J\"arvel\"a\footnote{European Space Astronomy Centre (ESAC), European Space Agency (ESA), 28692 Villanueva de la Ca\~{n}ada, Spain.}, \and P. Romano$^{*}$, \and S. Vercellone$^{*}$, \and E. Dalla Bont\`a$^{**}$}

\date{October 5, 2021}

\maketitle

\begin{abstract}
We are compiling a new list of gamma-ray jetted active galactic nuclei (AGN), starting from the fourth catalog of point sources of the \emph{Fermi} Large Area Telescope (LAT). Our aim is to prepare a list of jetted AGN with known redshifts and classifications to be used to calibrate jet power. We searched in the available literature for all the published optical spectra and multiwavelength studies useful to characterize the sources. We found new, missed, or even forgotten information leading to a substantial change in the redshift values and classification of many sources. We present here the preliminary results of this analysis and some statistics based on the gamma-ray sources with right ascension within the interval $0^{\rm h}-12^{\rm h}$ (J2000). Although flat-spectrum radio quasars and BL Lac objects are still the dominant populations, there is a significant increase in the number of other objects, such as misaligned AGN, narrow-line Seyfert 1 galaxies, and Seyfert galaxies. We also introduced two new classes of objects: changing-look AGN and ambiguous sources. About one third of the sources remain unclassified.
\end{abstract}

\section{Introduction: Why This Work?}
There are many ways to estimate the power emitted by a relativistic jet, but there are also serious problems of consistency among the various methods (e.g., \cite{PJANKA}). In 2019, we started a program aimed at calibrating the main methods to estimate the jet power and presented some preliminary results \cite{FOSCHINI19}. The need to compile a suitable list of gamma-ray emitting jetted active galactic nuclei (AGN) soon became apparent as the gamma-ray luminosity is a reliable proxy of the radiative jet power \cite{SIKORA,MARASCHI}. The obvious starting point was the fourth catalog of point sources of the {\em Fermi} Large Area Telescope (LAT) (4FGL, \cite{4FGL}) and the derived catalog of AGN (4LAC, \cite{4LAC}). 

However, the classification of the gamma-ray sources in the two catalogs is not exactly the same (see Section~\ref{stat}). Therefore, we decided to start a careful check. The classification is important in order to apply the most proper formula to calculate the jet power, which, in turn, depends on the main radiative process driving the gamma-ray emission. External Compton is dominant in objects with strong optical emission lines, such as flat-spectrum radio quasars (FSRQs) and narrow-line Seyfert 1 galaxies (NLS1s), while BL Lac Objects---characterized by optical spectra with weak or no emission lines---are powered by the synchrotron self-Compton.

 The classification is also important to understand when the jet emission is significantly contaminated by gamma rays from starburst activity as  happens in nearby Seyfert galaxies. Another important problem is to divide between beamed (the jet viewed at small angles) and unbeamed (large viewing angles), to understand how much special relativity can boost the jet power. 

To calculate the jet power, we also need to know a reliable redshift, where reliable means a spectroscopic redshift. Photometric redshifts are often found in the literature; however, in the case of jetted AGN, the strong variability of the jet emission can significantly affect the results. Therefore, we also need to carefully check the redshift values to keep only those based on optical spectroscopy. We also kept values estimated from the imaging of the host galaxy, although with a caveat.

To summarize, our philosophy was to select a sample of sources with redshift and classification that were as  reliable as possible, given the best information available.~It is better to have a relatively small, but reliable, sample, rather than a large one  with significant~uncertainties.

\section{Sample Selection and Procedure}
We started from the 4FGL Data Release 2 (DR2, Revision 5, 17 December 2020), which contains 5788 gamma-ray point sources\footnote{The FITS file can be downloaded here: \url{https://fermi.gsfc.nasa.gov/ssc/data/access/lat/10yr_catalog/} (accessed on October 4, 2021).}. We kept all the extragalactic sources (except for starburst and normal galaxies) plus the sources with counterparts of unknown nature (unk class), and the partially identified sources (bcu and blazar candidates of uncertain type). We did not consider any of the Galactic sources (pulsars, binaries, supernova remnants, star forming regions, etc.), the unassociated gamma-ray sources (i.e., without any counterpart), and the association with counterparts with a probability smaller than 80\%, which is also the threshold set by the {\em Fermi} LAT Collaboration to define an association. We also applied a cut to remove the sources at low Galactic latitudes ($|b|<10^{\circ}$), to take into account the issues related to the strong diffuse gamma-ray emission along the Galactic plane. The final sample consisted of $2982$ gamma-ray point sources associated with a counterpart. 

We carefully searched for the redshift of the counterparts both in the available literature and in public databases, such as the Sloan Digital Sky Survey (SDSS DR16\footnote{\url{http://skyserver.sdss.org/DR16/en/home.aspx} (accessed on October 4, 2021).}) and Large Sky Area Multi-Object Fiber Spectroscopic Telescope (LAMOST DR6V2\footnote{\url{http://dr6.lamost.org/v2/} (accessed on October 4, 2021).}). For the literature, we referred to the public databases Set of Identifications, Measurements and Bibliography for Astronomical Data (SIMBAD\footnote{\url{http://simbad.u-strasbg.fr/simbad/} (accessed on October 4, 2021).}), NASA/IPAC Extragalactic Database (NED\footnote{\url{http://ned.ipac.caltech.edu/} (accessed on October 4, 2021).}), and SAO/NASA Astrophysics Data System (ADS\footnote{\url{https://ui.adsabs.harvard.edu/} (accessed on October 4, 2021).}). 

We proceeded as follows: we took for granted the association proposed by the {\em Fermi} LAT Collaboration and searched for the counterpart in both SIMBAD and NED. For example, the first source in our sample is 4FGL~J$0001.2+4741$, which is associated with B3~$2358+474$. Therefore, we inserted B3~$2358+474$ in both SIMBAD and NED to search for published papers dealing with its identification and classification based on the optical and radio spectra, radio morphology, and optical imaging of the host galaxy. 

In this specific case, SIMBAD has only two publications, while NED has fourteen, but all of them are generic catalogs, and none of them contains the information we need. We searched also in SDSS and LAMOST, but there were no entries. The 4FGL classifies B3~$2358+474$ as bcu, and we  were not able to find either data or information suitable to identify this object. However, we adopted a more generic unclassified (UNCL) class (see the next section for an explanation of our adopted classes). 

Another example, the second source in our sample, which is 4FGL~J$0001.2-0747$ =~PMN~J$0001-0746$. Simbad reports 38 references, while NED has 33. No optical spectra were found in either the SDSS or LAMOST databases. Searching in the literature, we found only one useful reference \cite{SHAW13}, but the reported spectrum was featureless. Therefore, we confirmed the BL Lac Object classification of 4FGL, but without measured redshift.

In a further example, the third source: 4FGL~J$0001.5+2113=$~TXS~$2358+209$, Simbad has 16 references, while NED has 29. Both databases reported $z=1.106$ measured by \cite{FALCO}; however, searching in the available literature, we found that, some years later, this value was changed to $z=0.439$ by the same authors on the basis of a better spectrum \cite{MUNOZ}. The publicly available spectrum on the SDSS\footnote{\url{http://skyserver.sdss.org/DR16//en/tools/explore/summary.aspx?ra=0.384875&dec=21.226739} (accessed on October 4, 2021).} confirms the latter measurement. The spectrum also confirms the classification as flat-spectrum radio quasar set by 4FGL. This example also clearly shows why researchers should be wary of just downloading AGN redshifts from NED or SIMBAD without carefully vetting them, as many AGN have multiple measurements, and the NED/SIMBAD values may not reflect the most recent updates.

There are other cases where we found different redshift values for the same source. There are many reasons: wrong identifications based on weak features or artifacts, lower limits taken as measured values, and mere transcription errors from the paper to the online database. We would like to underline that the values of redshift and the revised classification were mostly taken from published works. We considered a redshift value from automatic pipelines of public databases (SDSS, LAMOST) only in the cases where it was evident. We did not perform any new observations or data analysis, because our aim is not to build a catalog but rather to extract a suitable list of gamma-ray emitting jetted AGN. 

\section{Classification}
The 4FGL and 4LAC adopted the following classes (acronyms in all capital letters indicate a firm identification, i.e., confirmed by coordinated multiwavelength variability)~\cite{4FGL,4LAC}:

\begin{itemize}
\item BLL/bll (BL Lac Object);
\item FSRQ/fsrq (flat-spectrum radio quasar); 
\item RDG/rdg (radio galaxy); 
\item AGN/agn (non-blazar active galaxy); 
\item SSRQ/ssrq (steep-spectrum radio quasar); 
\item CSS/css (compact steep spectrum radio source); 
\item BCU/bcu (blazar candidate of uncertain type); 
\item NLSY1/nlsy1 (narrow-line Seyfert 1 galaxy);
\item SEY/sey (Seyfert galaxy); and
\item UNK/unk (sources associated with counterparts of unknown nature). 
\end{itemize} 

We have adopted another classification scheme: 

\begin{itemize}
\item BLLAC: BL Lac Object;
\item FSRQ: flat-spectrum radio quasar;
\item NLS1: Narrow-Line Seyfert 1 Galaxy;
\item SEY: Seyfert galaxies (type 1, 2, or intermediate);
\item MIS: misaligned AGN, i.e., any of the above classes with the jet viewed at large angles;
\item CLAGN: changing-look AGN, when there are optical spectra at different epochs showing radical changes, such as from a featureless continuum to strong emission lines, thus indicating a change in the accretion history; a CLAGN must not be confused with a jetted AGN where the usual jet activity can hide or reveal some weak lines or features (for example, a BL Lac Object holds its classification when the jet activity overwhelms the weak lines and the Ca H\&K break, making the optical spectrum a featureless continuum);
\item AMB: ambiguous AGN, when the available information are contradictory or not sufficient for a reliable classification, or even for an educated guess; and
\item UNCL: unclassified, no optical spectrum available, no radio information useful to infer any misalignment. 
\end{itemize}

The main differences concern misaligned AGN and unclassified sources. We preferred a generic class UNCL, which includes the 4FGL classes UNK/unk and BCU/bcu. Regarding the latter, the blazar candidate of uncertain type (BCU/bcu) class is based on the localization in certain areas of some infrared photometric plots (e.g., \cite{DABRUSCO}). However, there are many reasons for having certain infrared colors and not all are related to a jet (for example, see the discussion in \cite{CACCIANIGA}). 

In addition, even a change in the jet activity can determine a change in the location of the photometric plots, drifting into spaces where different physical mechanisms are supposed to be the drivers of the electromagnetic emission. 
Last, but not least, it is already well known that the probability for a high-energy gamma-ray source to be a jetted AGN is greater than for any other type of source. Therefore, saying that a certain unknown gamma-ray source, particularly if located at high Galactic latitude, might be a blazar of uncertain type is an almost redundant information\footnote{It is worth mentioning that the definition of BCU/bcu has been extended in the 4LAC, by including the reference to a flat radio spectrum and/or the typical double-humped spectral energy distribution \cite{4LAC}.}. 

The most critical classification is that of misaligned AGN. There is no direct and clear-cut method to measure the jet viewing angle. All the methods are indirect (e.g., a Ca K\&H break in the optical spectrum, obscuration or not in the optical spectrum, radio core vs. extended emission, radio core brightness temperature, and radio spectral index), and suffer drawbacks (see the detailed discussion in \cite{URRYPAD}). For example, it is known that the Doppler factor $\delta$ depends on the bulk Lorentz factor $\Gamma=1/\sqrt{1-\beta^2}$ (where $\beta$ is the bulk speed in units of the speed of light in vacuum $c$) and the viewing angle $\theta$, according to the well-known equation:
\begin{equation}
\delta = \frac{1}{\Gamma\sqrt{1-\beta \cos \theta}}
\end{equation}

There is some degeneracy as shown in Figure~20 of \cite{URRYPAD}: $\delta=1$ can be obtained both with $\theta$$\sim$$20^{\circ}$ and $\Gamma=15$, and with $\theta$$\sim$$60^{\circ}$, and $\Gamma=2$. This means that all the methods based on the difference between beamed and unbeamed are subject to some degeneracy. Many authors set a threshold angle on the basis of different methods: for example, \mbox{Homan et al. \cite{HOMAN}} studied the brightness temperature and the apparent speed of the jet components ($\beta_{\rm app}$) and found that most of beamed jetted AGN were within $\theta \lesssim 15^{\circ}$ (see Figure~7 in \cite{HOMAN}), which was more or less confirmed also by \cite{JORSTAD}; 

Barthel \cite{BARTHEL} found a much larger value $\theta$$\sim$$44.4^{\circ}$ by using a statistical approach based on the optical properties. Given these uncertainties, the most conservative approach is to refer to published papers, where the geometry of the source was studied in detail by using different and complementary methods. This way is not exempt from drawbacks, as it depends on the different (or not) epochs of the observations, the intrinsic variability of the source, its distance from the Earth, the performance of the adopted instruments.  This is the best compromise given the available information.  

Our MIS class includes the 4FGL classes RDG/rdg, SSRQ/ssrq, and CSS/css: our choice is dictated by the simple need to separate beamed from unbeamed sources. Further details in the classification of these objects are beyond our aims. Nevertheless, we left in the Notes some additional information regarding the type of misaligned AGN (Fanaroff-Riley type 0, I, II, Broad-Line Radio Galaxy, Narrow-Line Radio Galaxy, etc.). 

The AMB class contains those jetted AGN with observational characteristics on the border between other well-defined classes, such as a BL Lac Object and a FRI radiogalaxy. The reasons can be the lack of data to break the degeneracy of the viewing angle or the intrinsic variability of the source or when the host galaxy is dominating the optical spectrum~\cite{BALMAVERDE}. The possibility of real hybrid objects cannot be ruled out: one can think, for example, about the recent discovery of a radio galaxy, PBC~J$2333.9-2343$, which underwent some cataclysmic event, and its jet is now aligned toward the Earth \cite{HERNANDEZ}\footnote{This specific object is taken as an example, but currently no gamma-ray emission has been detected.}. 

The classical definition of Narrow-Line Seyfert 1 Galaxy (NLS1) is based on the following quantities \cite{OSTERBROCK,GOODRICH}: (i) FWHM(H$\beta$) $\lesssim 2000$~km/s; (ii) [OIII]/H$\beta < 3$; and (iii) FeII bumps. The most recent research suggested that the main indicator of a NLS1 could simply be a narrow permitted H$\beta$, with $1000 \lesssim $ FWHM $\lesssim 2000$~km/s, while FeII bumps might not be so constraining \cite{CRACCO}. Therefore, when a measurement of the FWHM of the broad permitted H$\beta$ is available and within the defined range, we reclassify the source as NLS1. These sources were previously classified as FSRQ/fsrq, and, since jetted NLS1s are known to be the low-luminosity/-mass tail of the FSRQs distribution \cite{ABDO09,FOSCHINI15,BERTON16}, it might seem a useless distinction. 

This is due to the adoption of a classification based on an observed property that is  easy to measure: the equivalent width ($EW$) of the optical emission lines, with a threshold of 5~\AA~(FSRQ and NLS1, strong lines, $EW>5$~\AA; BLLAC, weak or no emission lines, $EW<5$~\AA). 
However, if we adopt a physical classification (e.g., see \cite{FOSCHINI17}, Figure~2), then FSRQs would be High-Mass Fast-Cooling (HMFC), while NLS1s would be Low-Mass Fast-Cooling (LMFC), and the difference would be clearer. 

This is more evident in the $L_{\rm disk}-P_{\rm jet}$ plane, where NLS1s make a branch different from FSRQs because of their relatively small mass of the central black hole and the high accretion luminosity \cite{FOSCHINI17}. Therefore, it is worth keeping this distinction even if, for the sake of simplicity, we still adopt an observation-based classification. 

We also decided to keep NLS1s separated from generic Seyfert galaxies (SEY class: type 1, type 2, or intermediate), since, in the latter cases, the physics driving the gamma-ray emission can be quite different, as the result of both a relativistic jet and star-burst activity~\cite{LENAIN,HAYASHIDA,GUO}. 

With the term changing-look AGN (CLAGN), we refer to those sources that can change their classification (e.g., from BLLAC to FSRQ or vice versa) after some dramatic event, such as a change in the accretion or in the jet activity \cite{ULRICH,FOSCHINI11,GHISELLINI,RUAN,HERAZO,MISHRA}. Sometimes the jet activity can determine changes in the observational appearance but not a change in classification: for example, the decrease/increase of equivalent width of the emission lines when the continuum increases/decreases because of the jet activity \cite{CORBETT,FOSCHINI12,BERTON18,BERTON21} or a shift of the synchrotron peak \cite{PIAN99,FOSCHINI08}. A word of caution must be set down when comparing optical spectra with very different signal-to-noise ratio (S/N): sometimes, the lack of emission lines or other features might be simply due to a combination of weak lines and low S/N spectra, not to an intrinsic variability of the AGN (e.g., \cite{GLIOZZI}).

We might set a list of high and low confidence classification, but also this option suffers drawbacks as pointed out by the presence of the newly established CLAGN class. This is really the tip of the iceberg. CLAGN shows that cosmic objects can significantly change their classification on human time scales. This label clearly depends on how much the object was observed in the past. For most of the checked sources, we found just one optical spectrum; therefore, it is easy to speculate about the possibility of some not observed change. The opposite case---a very well known source, observed for decades, such as BL Lac---is not a guarantee of a fixed classification (e.g., \cite{VERMEULEN}).  

The concluding words of wisdom are that any observation-based classification must be taken {\em cum grano salis}, i.e., must not be taken literally. It always reflects the available data, the epochs of observation, and the instrumental performance. Most importantly, time must be taken into account: any classification is a time-fixed scheme, while cosmic objects change their appearance in time but not their intrinsic physical nature \cite{LIVIO}. 

Therefore, this revision has to be intended, not as a tedious control, but as a step forward in understanding the nature of these sources. Further studies might require another change in the observation-based classification. This also emphasizes the need of a physics-based classification, as noted above for NLS1s, which would be more stable. Clearly, this type of classification can be done only after detailed studies and observations, and this is not the present case. 

Moreover, we also underline that our aim is not to generate a new catalog and/or to compete with the existing ones. This revision was made in the context of preparing a sample of gamma-ray emitting jetted AGN to be used for the calibration of the jet power. However, the information presented here are a useful spin off of the calibration project that we believe will be of use to the community. We hope it will be a starting point and an inspiration for many forthcoming studies.

\section{Preliminary Statistics}
\label{stat}
At the time of writing (20 August 2021), we have checked all the sources within the right ascension interval $0^{\rm h}-12^{\rm h}$ (J2000), corresponding to 1559/2982 sources ($\sim$$52$\%, \mbox{Figure~\ref{display}}). The full list of sources is available in the Appendix~A\footnote{The full list is available on the \href{https://www.mdpi.com/2218-1997/7/10/372}{published article}.}. For the sake of homogeneity, the redshifts were indicated with three significant digits and are the latest available, although the bibliographic reference reported the first consistent measured value (this means that there could be some slight difference in the latest significant digit between the value in the reference and the value in the table, which refers to the latest---and hopefully---best measurement).

\begin{figure}[h]
\centering
\includegraphics[scale=0.7]{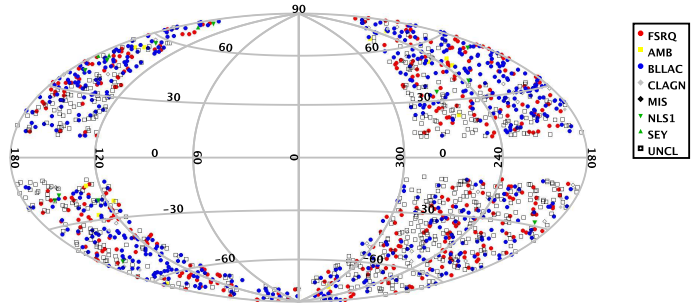}
\caption{Distribution in the sky (Galactic coordinates) of the checked sources. \label{display}}
\end{figure}   

The original classification of these 1559 sources in the 4FGL was distributed as follows: 

\begin{itemize}
\item 1 AGN, 3 agn;
\item 15 BLL, 598 bll;
\item 16 FSRQ, 353 fsrq;
\item 2 NLSY1, 1 nlsy1;
\item 3 RDG, 16 rdg;
\item 2 css;
\item 2 ssrq; and
\item 547 bcu.
\end{itemize}

After our check, here are some preliminary revised statistics on this first set of sources:

\begin{itemize}
\item 554/1559 BLLAC (35.5\%); 
\item 370/1559 FSRQ (23.7\%);
\item 12/1559 NLS1 (0.8\%);
\item 4/1559 SEY (0.3\%);
\item 39/1559 MIS (2.5\%);
\item 11/1559 CLAGN (0.7\%);
\item 14/1559 AMB (0.9\%); and
\item 555/1559 UNCL (35.6\%).
\end{itemize}

\begin{figure}[h]
\centering
\includegraphics[scale=0.7]{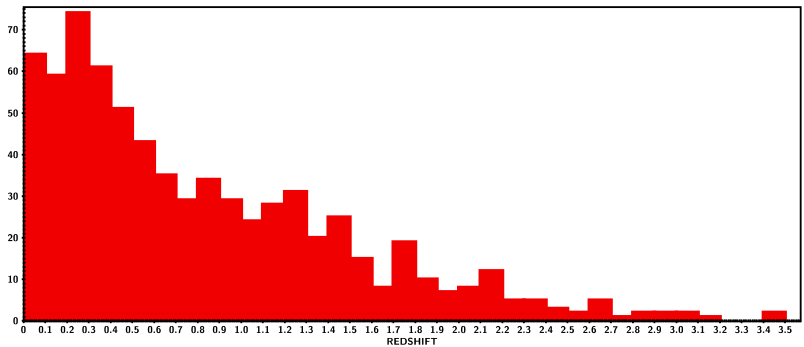}
\caption{Redshift distribution of the checked sources. \label{redshift}}
\end{figure}

Only 716 of the 1559 sources do have a redshift (46\%, Figure~\ref{redshift}). The values are spanning from $z=0.00828$ (4FGL~J$0958.3-2656$ = NGC~3078, AMB) to $z=3.45$ (4FGL~J$0337.8-1157$ = PKS~$335-122$, FSRQ; 4FGL~J$0833.4-0458$ = PMN~J$0833-0454$, FSRQ). Not all the sources classified as BLLAC  have a measured redshift: only 285/554 (51\%); the remaining have featureless or noisy spectra. The farthest BLLAC is 4FGL~J$0124.8-0625=$~PMN~J$0124-0624$ at $z=2.12$ \cite{SHAW13}.

A comparison with 4LAC \cite{4LAC} and 4FGL \cite{4FGL} statistics is displayed in Table~\ref{tab1}. The 4LAC clean sample is composed of all the extragalactic sources with $|b|>10^{\circ}$ and \texttt{FLAGS=0}. The \texttt{FLAGS=0} in the LAT catalog indicates the absence of systematic problems in the gamma-ray data analysis. However, a \texttt{FLAGS>0} does not necessarily mean the presence of artifacts: for example, Cen A has \texttt{FLAGS=512} because it has extended emission due to the jets. Therefore, in our sample, we included all the sources independently on the \texttt{FLAGS} value. 

The full 4LAC contains all the extragalactic gamma-ray sources, including also those on the Galactic plane. The 4FGL has all the gamma-ray sources detected by {\em Fermi} LAT, including the Galactic sources (pulsars, pulsar wind nebulae, supernova remnants, ...). We compared our sample also with the original subsample of the 4FGL (4FGL-O) that we used as starting point for our analysis. 

4LAC and the corresponding classes in the 4FGL are not exactly matching (4LAC full vs. 4FGL: FSRQ 655 vs. 694; BLLAC 1067 vs. 1131; $\sim$$6$\% difference in both cases\footnote{This difference is not completely explained by the fact that 4LAC contains only sources with {\tt FLAGS~=~0}, while 4FGL includes all the gamma-ray sources, independently of their {\tt FLAGS} value.}), indicating some changes/update/different opinions in the classification methods adopted by two different working groups of the same {\em Fermi}/LAT collaboration (the two papers have been submitted almost at the same epoch, with just a few months~difference). 

The most striking differences between our list and the original LAT catalogs refer to the secondary classes (MIS, NLS1, and SEY as well as the new classes CLAGN and AMB), while the main classes (BLLAC and FSRQ) have relatively small or no changes in percent. Although, as we have already noted, MIS is the most difficult class to be defined, and we  adopted a very conservative approach (the use of published papers), the number of sources in our sample is comparable with the numbers of the full catalogs. It is almost doubled in percent.  NLS1 is worth noting, as it is almost tripled in percent. Again, we referred only to published values of FWHM(H$\beta$), but the quick look of some yet unanalysed optical spectra of the SDSS suggests that some specific data analysis might reveal even more NLS1s. The SEY class is slowly emerging and deserves  attention. 

\begin{table}[h]
\caption{Comparison with statistics of 4LAC Clean samples (extragalactic sources with $|b|>10^{\circ}$ and \texttt{FLAGS = 0}), 4LAC full sample (all sky), and 4FGL (data from tables in \cite{4LAC,4FGL}). In the cases of 4LAC and 4FGL, the MIS class corresponds to the sum of the classes RDG/rdg, CSS/css, and SSRQ/ssrq. 4FGL-O refers to the original classification of the same dataset of the 4FGL.}
\begin{center}
\footnotesize
\begin{tabular}{lccccc}
\hline
\textbf{Class}		& \textbf{Present Work} & \textbf{4FGL-O}	& \textbf{4LAC-Clean} & \textbf{4LAC-Full} & \textbf{4FGL} \\
\hline
Nr. Src				& 1559					& 1559				& 2614			&  2863 			& 5064 \\
\hline
BLLAC & 554 (35.5\%)			& 613 (39.3\%) 		&	1027 (39.3\%) 	&  1067 (37.3\%) 	& 1131 (22.3\%) \\
FSRQ				& 370 (23.7\%)  		& 369 (23.7\%)		& 591 (22.6\%) 	&  655 (22.9\%)		& 694 (13.7\%) \\
NLS1				& 12 (0.8\%)     		& 3 (0.2\%)			& 9 (0.3\%)   		& 9 (0.3\%) & 9 (0.2\%) \\
SEY                 & 4 (0.3\%)     		& 0 (0\%)			& 0 (0.0\%)   		& 0 (0.0\%) & 1 (0.02\%) \\
MIS					& 39 (2.5\%)    		& 23 (1.5\%)		& 38 (1.4\%)   		& 45 (1.6\%) & 49 (1.0\%) \\
\hline
\end{tabular}
\normalsize
\end{center}
\label{tab1}
\end{table}%

\section{Final Remarks}
Starting from the fourth catalog of gamma-ray sources detected by {\em Fermi}/LAT, we are building a list of gamma-ray emitting jetted AGN to be used in a project of calibration of  jet power. We presented here some preliminary statistics on the first half of the selected sources (RA $0^{\rm h}-12^{\rm h}$, J2000), which shows the emergence of an unexpected treasure of novelties that are awaiting study. Even with all the required caveats, it is now evident that an observation-based classification of jetted AGN cannot be a time-fixed scheme, but it should be regarded in a more dynamical way, to take into account not only the apparent changes of the cosmic sources but also the improvements in the observing technology as well as the proper use of the archives containing published articles and unpublished data. 

In particular, we would like to emphasize the importance of looking at the published papers, because they contain information that cannot be extracted with a simple cross-match between~catalogs. 

\vspace{6pt}

\section*{Acknowledgments}
\footnotesize
LF thanks Alessandro Caccianiga for a useful and valuable exchange of ideas on the classification of misaligned AGN. Funding for the Sloan Digital Sky Survey (SDSS) was provided by the Alfred P. Sloan Foundation, the Participating Institutions, the National Aeronautics and Space Administration, the National Science Foundation, the U.S. Department of Energy, the Japanese Monbukagakusho, and the Max Planck Society. The SDSS Web site is \url{http://www.sdss.org/}. The SDSS is managed by the Astrophysical Research Consortium (ARC) for the Participating Institutions. The Participating Institutions are the University of Chicago, Fermilab, the Institute for Advanced Study, the Japan Participation Group, The Johns Hopkins University, Los Alamos National Laboratory, the Max-Planck-Institute for Astronomy (MPIA), the Max-Planck-Institute for Astrophysics (MPA), New Mexico State University, University of Pittsburgh, Princeton University, the United States Naval Observatory, and the University of Washington. Guoshoujing Telescope (the Large Sky Area Multi-Object Fiber Spectroscopic Telescope LAMOST) is a National Major Scientific Project built by the Chinese Academy of Sciences. Funding for the project has been provided by the National Development and Reform Commission. LAMOST is operated and managed by the National Astronomical Observatories, Chinese Academy of Sciences. This research  made use of the SIMBAD database, operated at CDS, Strasbourg, France (2000, A\&AS, 143, 9, ``The SIMBAD astronomical database'', Wenger et al.). This research  made use of the NASA/IPAC Extragalactic Database (NED), which is funded by the National Aeronautics and Space Administration and operated by the California Institute of Technology. This research  made use of NASA’s Astrophysics Data System Bibliographic Services.

\end{document}